\documentclass[letterpaper]{article}
\usepackage{verbatim}
\usepackage{amssymb,enumerate}
\usepackage{graphicx}

\begin{document}

\title{Modeling and Verification of Agent based Adaptive Traffic Signal using Symbolic Model Verifier}
\author{Vivek Vishal, Sagar Gugwad and Sanjay Singh\thanks {Sanjay Singh is with the Department of Information and Communication Technology, Manipal Institute of Technology, Manipal University, Manipal-576104, INDIA, E-mail: sanjay.singh@manipal.edu}}

\maketitle
\begin{abstract}
This paper addresses the issue of modeling and verification of a Multi Agent System (MAS) scenario. We have considered an agent based adaptive traffic signal system. The system monitors the smooth flow of traffic at intersection of two road segment. After describing how the adaptive traffic signal system can efficiently be used and showing its advantages over traffic signals with predetermined periods, we have shown how we can transform this scenario into Finite State Machine (FSM). Once the system is transformed into a FSM, we have verified the specifications specified in Computational Tree Logic(CTL) using NuSMV as a model checking tool. Simulation results obtained from NuSMV showed us whether the system satisfied the specifications or not. It has also showed us the state where the system specification does not hold. Using which we traced back our system to find the source, leading to the specification violation. Finally, we again verified the modified system with NuSMV for its specifications. 
\end{abstract}

\section{Introduction}
\label{sec1}
A significant amount of work has been done in the field of artificial intelligence. It has resulted in the development of different types of logics, semantics and tools but very few work focuses on its verification. Rao \cite{Rao} introduced the agent-oriented programming language AgentSpeak(L). It tried to bridge up the gap between the theoretical perspective and practical design. It served as one of the major breakthrough in the agent technology and gave rise to many research activities in this direction.
\par
 All of the early work done concentrated in the development of different variations of this language to support their domain and related tools to build applications. Bordini \cite {bor} came up with his AgentSpeak(F), a finite state version of AgentSpeak(L), to support model checking of applications developed in AgentSpeak(L). 
 \par
In this paper we have considered a scenario of adaptive traffic signal across the intersection of two roads (any other scenario is just a special case of it). Duration of green light vary based on the length of traffic at a particular junction of the road in such a way that each vehicle gets a fair opportunity to cross the intersection. Modeling of this scenario has been shown and the required properties of the model has been verified using NuSMV \cite{nu}\cite{nus}.

\par
This paper is organized as follows. Section \ref{sec:formal} gives an introduction to current traffic scenario which is based on predetermined traffic signal periods. Section \ref{sec:model} discusses the detail design of agent based adaptive traffic monitoring system. Section \ref{sec:spec} discusses about model checking and the properties of the agent based adaptive traffic signal, specified using CTL. Section \ref{sec:result} explains about the verification results obtained. Finally section \ref{sec:conclusion} concludes this paper.

\section{Current Traffic Scenario}
\label{sec:formal}

Traffic jams are becoming a daily routine on roads especially in cities with large population. The main reason for this is the increase in the number of on-road vehicles Vs. limited road area. One solution of this problem can be the construction of new roads or widening of already existing roads. This is however not possible in most of the cases due to the scarcity of land and other issues. In that case we are only left out with the option of efficient flow-management of the on-road vehicles particularly  at intersection point of two roads. \par
	Today the only flow-management available is through traffic signals. Traffic light systems are built to control the traffic flows at the intersections to ensure the fluency of traffic flow within the traffic network. In this paper we tried to convert this "flow-management aid" to an "effective flow-management aid" for controlling the traffic in efficient  way so that we can manage the available traffic to the limited road space. Our aim here is to ensure a fair chance to each vehicle waiting for a green signal.\par
	\begin{figure}[bpht!]
	\centering
		\includegraphics[height=8cm,width=12cm]{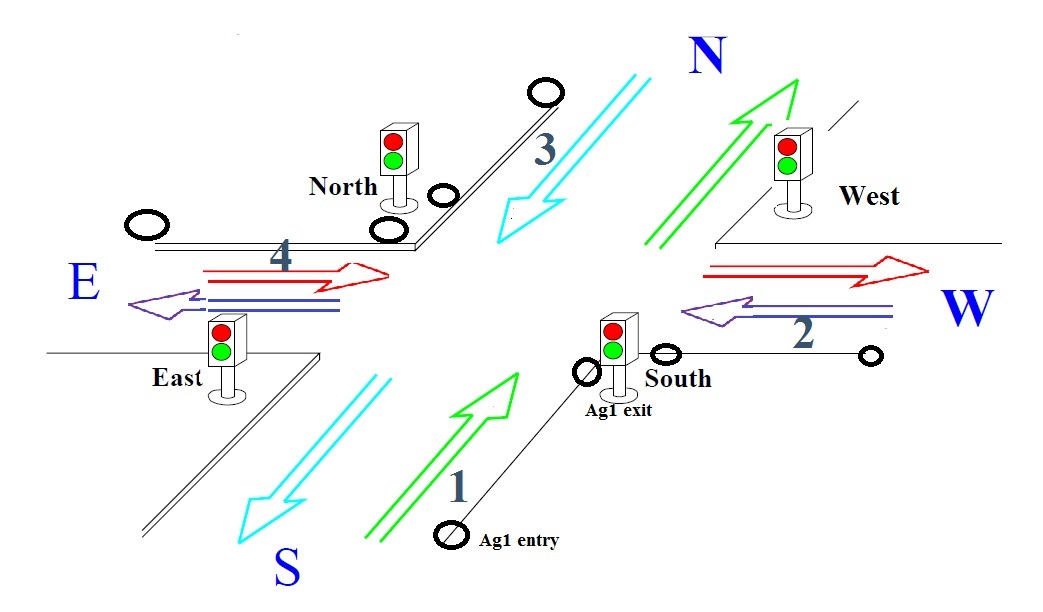}
		\caption{Adaptive Traffic Scenario}
		\label{fig:f1}	
\end{figure}
	Nowadays, most of the traffic light control systems are still using predetermined setting. Consider the road intersection scenario as shown in the Fig.\ref{fig:f1}. Suppose the lanes 1 and 2 are more busy than lanes 3 and 4. Due to the predetermined settings of the traffic lights, the duration for  green signal for each lane is fixed. As a result whenever there are no vehicles on lane 3 and 4 their lights will be green for the entire fixed interval. Even if there are more vehicles on lane 1 and 2 ,they will have to wait unnecessarily. This results in the increase of vehicle queue in lanes 1 and 2. This process continues as the cycle proceeds leading to traffic jam. This is one of the major drawbacks of using traffic signals with predetermined settings.\par
	
	Many works has been done in the past in order to overcome the problem caused by the usage of predetermined traffic signals. Teo and Chin \cite{Teo} has extended the concept of Genetic Algorithm for the optimization of  the Traffic Flow Control. Longer green time will pass through more vehicles, but it will increase the cycle time at the same time which causes more vehicles to accumulate at the intersection during the waiting time. Using Genetic Algorithm the optimal value of the green time of the signal is determined. They have predicted the duration of current green time on the basis of its wait time that is duration for which the other lanes were green. Since this approach is based on prediction and not on any current real time data, the whole system cannot be considered as reliable as far as average waiting time for each vehicle is concerned.\par
	Zhou et al \cite{Zhou} has proposed an algorithm which takes care of almost all the considerable parameters such as traffic volume, waiting time, vehicle density, etc., to determine green light sequence and the optimal green light duration. One drawback in this approach is that they have considered 12 possible configuration of green light, which is indeed not required and is a system overhead.\par
	Cheng \cite{Cheng} has also proposed a similar approach by applying an improved adaptive PI algorithm to calculate the weight of each state, which is the key to determine the next state of a traffic light. If the current state is identical to the expected one, green light period is extended for the current state, otherwise the expected state is switched on, and its green light period is launched. However as in the previous paper \cite{Zhou} they have also considered 12 possible configuration of green light and suffers from same drawback. 
\par
A number of literature \cite{Hir}\cite{Kul}\cite{Cha} have addressed the issue of adaptive traffic signal via wireless sensor networks,  machine learning, and other approaches.

\section{Agent Based Adaptive Traffic Monitoring System}
\label{sec:model}

The aim of the agent based traffic monitoring model is to provide an efficient flow of traffic across intersections of two or more roads. In our scenario we have considered an intersection of two road segments. Each road segment has two lanes. Though we have tried to take a generalized scenario, the real world road intersections can be of higher dimension. This model can easily be fitted to model even those scenarios with a little modification.
\par
In this approach, instead of taking twelve possible combinations of green light we have taken only four possible combination of green light as shown in the Fig.\ref{fig:f1}. Each one of the four combinations will get a chance in a weighted round-robin fashion. Now the question is how we can determine the weight for the round-robin that is for how much duration that green light will be ON. The metric for this weight is the queue length at the time when the green light is about to turn ON.  Before  going into the detailed design let us define some terminologies used:
 
%\begin{figure}[bpht!]
%	\centering
%		\includegraphics[scale = 0.5]{s1}
%		\caption{Adaptive Traffic Scenario}
%		\label{fig:f1}	
%\end{figure}
% 
 
\begin{itemize}
 \item $T_{thr}$: maximum duration for which any green light can be turned ON.
 \item $t_v$    :time required for any vehicle to cross the intersection. For simplicity we have assumed that it is constant and for modeling it in NuSMV we have taken it as 1. 
 \end{itemize}

Each incoming lane (towards the intersection) of a road segment has two monitoring agents which can be a sensor in real life scenario. Let Ag1Entry and Ag1Exit respectively be that two agents for lane 1 for counting the number of incoming and outgoing vehicles from lane 1 as shown in Fig.\ref{fig:f1}. Ag1entry is placed somewhat far from the intersection, in such a way that its distance from the intersection point is at least $T_{thr} \times t_v$. By doing  so, we are ensuring that even if the vehicle queue exceeds the location point of the agent, the Agent is not making a count it. Ag1Exit Agent is situated at the intersection point. Similarly for all other three incoming lanes we have two monitoring agent for each lane, giving us a total of eight agents. There is yet another agent, Ag\_Master which determines the weight that is the duration for which a particular green light is ON.
\par
Now we will see how this Multi Agent System (MAS) works in a cooperative manner in order to provide an optimal weight for each green signal in each round. The Ag1Entry and Ag2Exit agents are very simple agents whose work is simply to pass a signal to the master agent, Ag\_Master as and when a vehicle passes through them. Now it is the master's job to determine the queue length of a particular lane just before it allocates a green signal to it.
\par
The master maintains a counter for each Entry/Exit Agents. Initially all the counters are initialized to zero. As and when it receives a signal from any  Entry/Exit agent it updates the counter of that particular agent by 1. Suppose that the current green signal is about to die, at that instance the master will determine the weight of the next (determined by round-robin) green signal by subtracting the number of exit counter from the entry counter for that particular lane. This subtraction gives us the number of vehicles in the queue in a particular lane at that particular instance of time. 

\begin{figure}[bpht!]
	\centering
		\includegraphics[height=8cm,width=14cm]{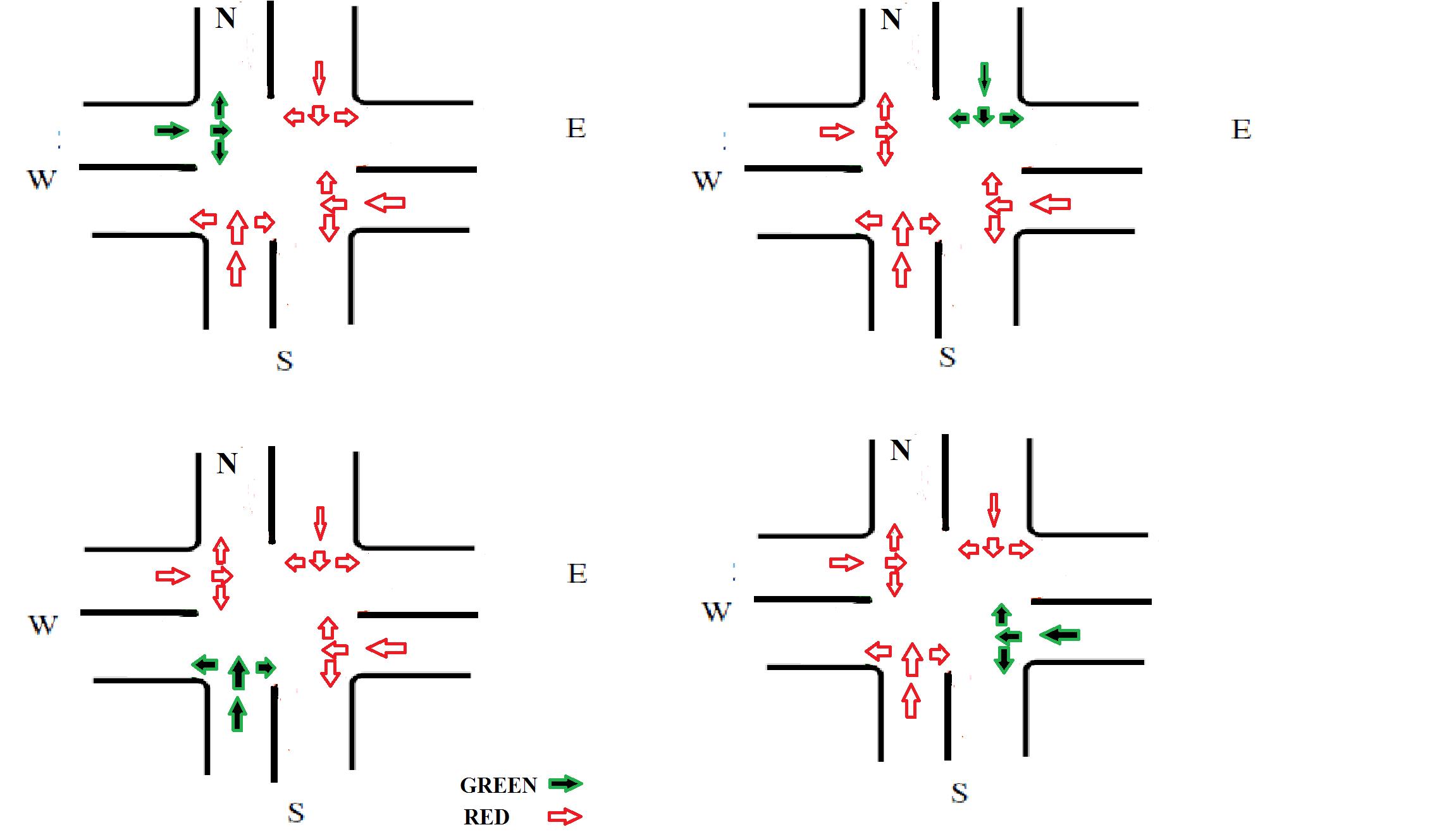}
		\caption{Weighted Round-Robin}
		\label{fig:f2}	
\end{figure}
Now the master agent know the number of vehicles (say n) in the particular lane at the time of its scheduling. It also knows the time required by a vehicle to cross the intersection that is $t_v$. Therefore, time required for all the vehicles to cross is $n\times t_v$. If this value is less than the $T_{thr}$, then a weight of $n\times t_v$ is allocated  to that green signal otherwise a weight of $T_{thr}$ is assigned to it.
\par
Here one question may arise that why do we need two agents when only one agent is sufficient? The answer is that as far as there is no traffic over head that is number of vehicles is less than the threshold value then this approach works fine. However, what about a situation when the number of vehicles are more than the $T_{thr}$. In such situations how the master agent come to know how many vehicles have passed and how many are still in queue. Under such scenario it is unable to predict the number of vehicles after one iteration in a particular lane. Therefore, at least two agents are required for discovering the correct number of vehicles in each lane.

\section{Model Checking}
\label{sec:spec}
Model checking \cite{huth} is a formal and automatic technique used to verify computational systems against given properties. They are applicable for finite state systems and they generally operate on system models and not on the actual system. The system is represented by a finite model $\mathcal{M}$ and the specification is represented by a formula $\phi$ using an appropriate logic. The verification method consists of computing whether the model $\mathcal{M}$ satisfies $\phi$ (i.e.$\mathcal{M} \models \phi$) or not (i.e.$\mathcal{M} \not \models \phi $). 
\par
The model checker either confirms that the properties hold or reports that they are violated. In the latter case, it provides a counter example: a run that violates the property. Such a run can provide valuable feedback and points to design errors. Model checking is based on temporal logic. The idea of temporal logic is that a formula is not statically true or false in a model, as it is in propositional and predicate logic. Instead, the models of temporal logic contain several states and a formula can be true in some states and false in others. Thus, the static notion of truth is replaced by a dynamic one, in which the formulas may change their truth values as the system evolves from state to state. In model checking, the models $\mathcal{M}$ are transition systems expressed in terms of FSM and the properties $\phi$ are formulas in temporal logic either LTL or CTL. To verify that a system satisfies a property, we must do three things:
\begin{enumerate}	 
\item Model the system using the description language of a model checker, arriving at
a model $\mathcal{M}$;
\item Code the property using the specification language of the model checker, resulting
in a temporal logic formula $\phi$;
\item Run the model checker with inputs $\mathcal{M}$ and $\phi$.
\end{enumerate}
\section{Finite State Machine For Agent Based Adaptive Traffic Signal System}
\label{sec:result}
In this section we will see how we can model our current traffic scenario into a Finite State System (FSM) and verify its properties against the specification. The FSM representation of traffic scenario F can be expressed as :\\
$ F:(Q,T,S_o,L)$. Where,
\begin{itemize}
\item Q : Set of States.\\
      The system can be in any one state at any moment of time. Each state denotes a particular instance of time. The number of states in this system is finite but is non-deterministically determined on the basis of number of vehicles in a particular lane at a particular moment of time.

\item T : Transition Function.\\
      A transition function determines how we can move from one state to another. In this scenario, upon every tick of clock we move from one state to another.

\item L : Labeling Function.\\
     It states which variables takes what value in which state. It should be noted that every variable declared in NuSMV is present in every state of the system.

\item $S_o$ : Starting state.\\ 
     It is state from where system start executing. It comes only once for the considered scenario.
\end{itemize}	

Readers having some knowledge of FSM might be wondering why we haven't shown the terminating state of the system. It is simply due to the reason that this system never terminates.   
\subsection{State Transition Diagram}
\begin{figure}
	\centering
		\includegraphics[scale =0.5]{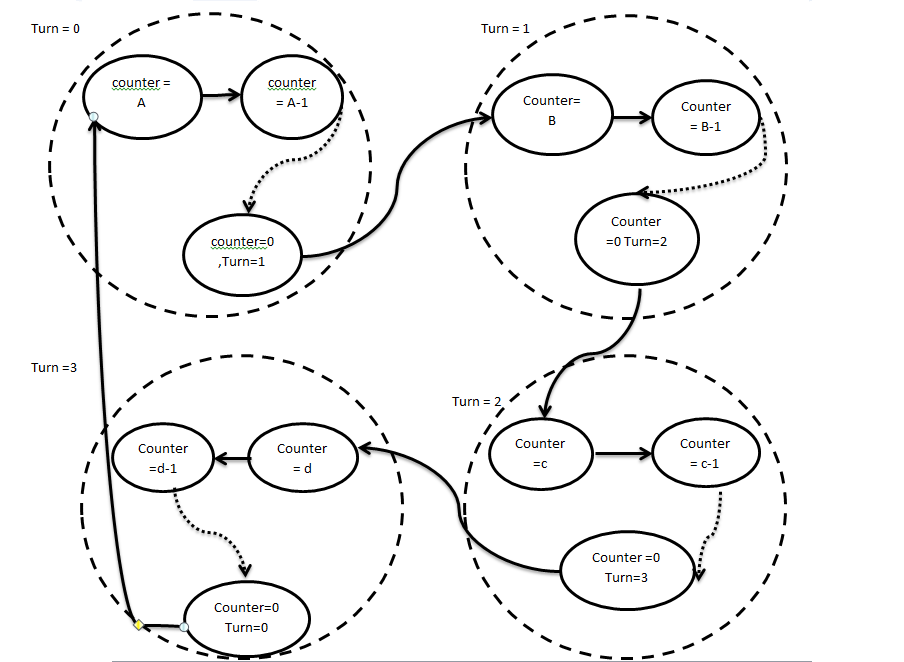}
		\caption{State Transition Diagram}
		\label{fig:modal}
\end{figure}
The state transition diagram of the adaptive traffic scenario is shown in Fig.\ref{fig:modal}. Initially, we assume that the system begins its execution from signal post labeled as NORTH in the Fig.\ref{fig:f1}. The system proceeds in a weighted round-robin fashion as NORTH, WEST, SOUTH, EAST, NORTH and so on. The duration of time allocated to each signal is directly proportional to the number of vehicles in that road segment. Mapping the Fig.\ref{fig:f1} to the modal in Fig.\ref{fig:modal}, we have,
\begin{enumerate}[(i)]
\item Turn = 0 corresponds to green light of signal post at NORTH. 
\item Turn = 1 corresponds to green light of signal post at WEST. 
\item Turn = 2 corresponds to green light of signal post at SOUTH. 
\item Turn = 3 corresponds to green light of signal post at EAST. 
\end{enumerate}

The master agent Ag\_Master at the time of expiry of green light of the previous signal allocates a counter value to the current green signal. The counter value depends not only on the number of vehicles but also on $T_{thr}$. Let the time duration calculated by master agent $Ag\_Master$ is $T_{cal}$. Now the counter value (CV) is calculated as \\ 
                   $CV = min\{T_{cal}, T_{thr}\}$\\ 
Once the counter value is determined, it keeps on decrementing, moving from one state to another till counter value becomes zero. Once the counter value becomes zero, the turn moves on to the next signal. It again computes the counter value for that particular signal and proceeds further in a round-robin fashion.
\par
Moving further, each state has wait time counter. The function of this wait time counter is to compute the waiting time of a green signal after it goes red and before it becomes green again. This measure gives us information regarding average waiting time and waiting time of vehicles in a particular road segment. Using wait time measurement we have checked the specification whether the maximum waiting time for any vehicle exceeds any particular predetermined value.
\subsection{Specification Verification using NuSMV}

NuSMV \cite{huth}\cite{nu}\cite{nus} stands for 'New Symbolic Model Verifier.' NuSMV is an Open Source product, is actively supported and has a substantial user community. NuSMV provides a language for describing the models we have been drawing as diagrams and it directly checks the validity of LTL (and also CTL) formulas on those models. NuSMV takes as input a text consisting of a program describing a model and some specifications expressed in terms of temporal logic formulas. It produces as output either the word \textit{true} if the specifications hold, or a trace showing why the specification is \textit{false} for the model represented by our program.
\par
In real world traffic scenario the value of $T_{thr}$ can be 180 seconds. Thus, any vehicle will have to wait for a maximum of $180 \times 3 = 540$ seconds. However, as a model is abstraction of real world application, we have scaled down $T_{thr}$ to 18. We can also simulate the system with value of $T_{thr}$ as 180 but doing so will be computing intensive process. If the required properties are satisfied in the model, then it will also be satisfied in real system. Therefore, we can carry on with our abstraction.  
\par
	In the agent based adaptive traffic monitoring system the desired properties of the system and their specification are :
	\begin{enumerate}[(1).]
	\item To verity that whether each signal post is getting a chance in a round-robin fashion.\\This property holds good if NORTH signal is followed by WEST signal, WEST signal by SOUTH signal, SOUTH signal by EAST signal,which is followed by NORTH. This property can be specified in CTL as :\\
SPEC AF (light0.counter = 0 $\rightarrow$ AX light1.colour =green)\\
SPEC AF (light1.counter = 0 $\rightarrow$ AX light2.colour =green)	\\						
SPEC AF (light2.counter = 0 $\rightarrow$ AX light3.colour =green)\\
SPEC AF (light3.counter = 0 $\rightarrow$ AX light0.colour =green)\\\\
In simple English whenever in any future light0.counter becomes zero then always the next state would be the state in which light1.colour = green. The other specifications can be analyzed in a similar way.

\item To verify that a signal post always gets an opportunity to turn green in some future state. The CTL spec in NuSMV is as follows:\\
SPEC AG(light0.colour =red $\rightarrow$ AF light0.colour =green)\\
SPEC AG(light1.colour =red $\rightarrow$ AF light1.colour =green)\\
SPEC AG(light2.colour =red $\rightarrow$ AF light2.colour =green)\\
SPEC AG(light3.colour =red $\rightarrow$ AF light3.colour =green)\\\\
The specification states that for any state if light0.colour = red then eventually in some future state it will become green.

\item To verify that maximum waiting time for a particular vehicle does not exceeds a predetermined value.The CTL spec would be:\\
SPEC AG (light0.wait $\le$ 54)\\
SPEC AG (light1.wait $\le$ 54)\\
SPEC AG (light2.wait $\le$ 54)\\
SPEC AG (light3.wait $\le$ 54)\\
The specification states that in every state the wait time for a particular signal should not be more than 54.
\end{enumerate}

\subsection{Observations and Results}
\begin{figure}[bpht!]
	\centering
	\includegraphics[height=12cm,width=12cm]{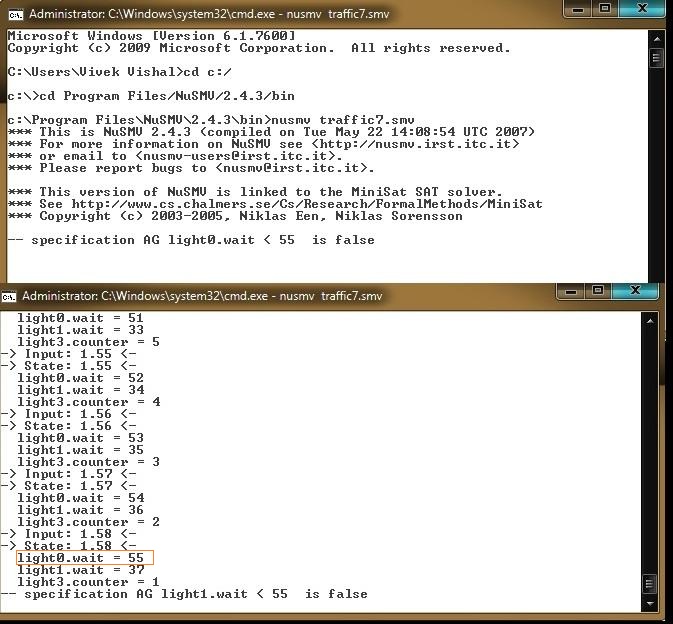}
	\caption{Simulation Result Before Correction}
	\label{fig:nu}	
\end{figure}

	While simulating the stated properties in NuSMV we found that the first two properties holds good in all situations but there is some contradiction with third property. The maximum waiting time for any green signal can be atmost $3 \times T_{thr}$. This situation can occur when the traffic is at its peak level in all the lanes. In such case time duration of each green signal will equal to $T_{thr}$. Therefore, in our model the maximum waiting time for any green signal is 54 $(3 \times 18$). However, the simulation results obtained from NuSMV as shown in Fig.\ref{fig:nu} shows that this specification does not hold. While tracing back the results, we have found that the programmer has taken the value counter from 0 to 18 which is actually 19 counts. Therefore, according to this value of counter the maximum waiting time is $19 \times 3$ which comes out to be 57.
		
\begin{figure}[bpht!]
	\centering
	\includegraphics[height=8cm,width=12cm]{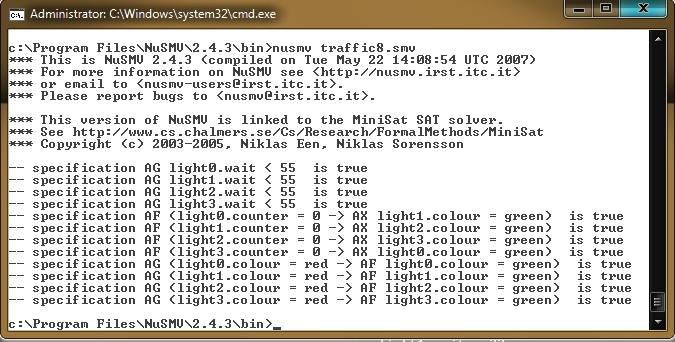}
	\caption{Simulation Result After Correction}
	\label{fig:nus}	
\end{figure}

Hence, if the system was deployed without any previous simulation each lane has to wait 3 seconds more than its actual wait time value. After correcting the value of counter from "0 to 18" to "0 to 17", we get the correct simulation result as shown in the Fig.\ref{fig:nus}.

\begin{figure}[bpht!]
	\centering
		\includegraphics[scale=0.8]{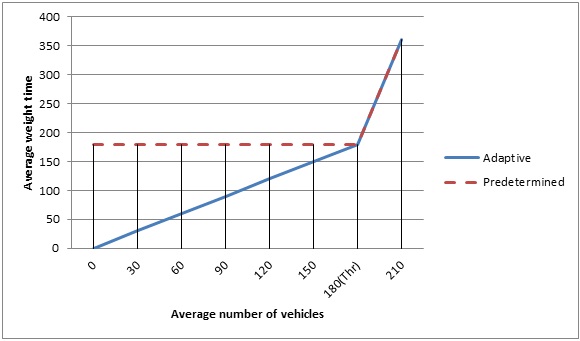}
		\caption{Comparison Graph}
		\label{fig:graph}
\end{figure}
\subsection{Performance Analysis}
Let us consider a real life traffic scenario in which $T_{thr}$ is 180 seconds and $t_v$ is 1 second. For predetermined traffic signal, let the value of periods be 180 seconds. As long as the number of vehicles in all the lanes are less than the threshold $T_{thr}$, Agent Based Adaptive Traffic Monitoring System will show higher performance with respect to predetermined traffic periods based systems. Once the traffic of all the lanes reaches its threshold $T_{thr}$, both systems will show identical performance as shown in the Fig.\ref{fig:graph}. Average waiting time for each vehicle in adaptive system is far less than the average waiting time of predetermined traffic system.
%\begin{figure}[bpht!]
%	\centering
%		\includegraphics[scale=0.8]{wait}
%		\caption{Comparison Graph}
%		\label{fig:graph}
%\end{figure}

\section{Conclusion}
\label{sec:conclusion}
There is a great advantage in being able to verify the correctness of computer systems, whether they are hardware, software, or a combination. As model checking considers each and every state, where a system can go in its entire lifetime, it can easily detect under what conditions the required properties of the system will not hold.\par
Results obtained from agent based adaptive traffic signal ensures a fair chance to cross an intersection for every vehicle in each lane based on their wait time. The system also takes care of the idle green light duration that is green light for a particular lane is turned ON even when their are no vehicles in that particular lane. The system has very effective results when the traffic is under a threshold value and reduces average waiting time significantly. When the traffic is above that threshold level, the performance of agent based adaptive traffic signal is same as that of predetermined.   
%\par
%There is a great advantage in being able to verify the correctness of computer systems, whether they are hardware, software, or a combination. As model checking considers each and every state, where a system can go in its entire lifetime, it can easily detect under what conditions the required properties of the system will not hold. We have also shown how  agent based adaptive traffic signal can be transformed into FSM and how we can verify its properties using NuSMV. 
\par
The work done in this paper regarding the effective traffic monitoring system has ignored some of the real world constrains for the sake of simplicity. This includes traffic lights for civilians crossing the roads on foot, providing some mechanism for giving pass to ambulances, VIPs etc. Our future work will be directed in the way of providing all real world constrains and verifying the system using formal method in NuSMV.

\bibliographystyle{IEEEtran}
\bibliography{myref}

\end{document}